\documentclass[superscriptaddress,twocolumn,aps,prl,floats,showpacs]{revtex4}
\usepackage{epsfig}% Include figure files
\usepackage{dcolumn}% Align table columns on decimal point
\usepackage{bm}% bold math
\begin{document}

\title{Glass Transition in a Two-Dimensional Electron System in Silicon in a
Parallel Magnetic Field}
\author{J.\ Jaroszy\'nski}
\altaffiliation[Also at ]{Institute of Physics, PAS, Warsaw, Poland}
\email{jaroszy@magnet.fsu.edu}
\affiliation{National High Magnetic Field Laboratory, Florida State
University, Tallahassee, Florida 32310}
\author{Dragana  Popovi\'c}
\affiliation{National High Magnetic Field Laboratory, Florida State
University, Tallahassee, Florida 32310}
\author{T.\ M.\ Klapwijk}
\affiliation{Department of Applied Physics, Delft University of
Technology, 2628 CJ Delft, The Netherlands}
\date{\today}

\begin{abstract}
Studies of low-frequency resistance noise show that the glassy freezing of the
two-dimensional electron system (2DES) in Si in the vicinity of the
metal-insulator transition (MIT) persists in parallel magnetic fields $B$ of
up to 9~T.  At low $B$, both the glass transition density $n_g$ and $n_c$, the
critical density for the MIT, increase with $B$ such that the width of the
metallic glass phase ($n_c<n_s<n_g$) increases with $B$. At higher $B$, where
the 2DES is spin polarized, $n_c$ and $n_g$ no longer depend on $B$.  Our
results demonstrate that charge, as opposed to spin, degrees of freedom are
responsible for glassy ordering of the 2DES near the MIT.
\end{abstract}


\maketitle

The fascinating strong correlation physics exhibited by
low-density two-dimensional (2D) electron and hole
systems~\cite{SAK2000} remains the subject of intensive research.
In the vicinity of the apparent metal-insulator transition (MIT),
in particular, both electron-electron interactions and disorder
appear to be equally important.  Their competition may lead to the
emergence of many metastable states and the resulting glassy
dynamics of electrons.  Recent experiments~\cite{SBPRL02,JJPRL02}
on a 2D electron system in Si have demonstrated such glassy
behavior, lending support to the theoretical proposals that
attempt to describe the 2D MIT as the melting of a
Coulomb~\cite{Thakur,Pastor}, Wigner~\cite{Sudip}, or
spin~\cite{Sachdev} glass.  Even though several features of the
data~\cite{SBPRL02,JJPRL02} are consistent with the model of
glassy behavior that occurs in the charge
sector~\cite{Pastor,Denis,Darko}, it is still an open question
whether charge or spin degrees of freedom are responsible for the
observed glass transition. Since a sufficiently strong magnetic
field is expected to destroy the spin glass
order~\cite{Binder,Sachdev}, experimental studies of glassy
dynamics in parallel magnetic fields $B$~\cite{comment1} should be
able to distinguish between the proposed models.  Here we present
such a study, which shows that the glass transition persists even
in $B$ such that the 2D system is spin polarized.  These results
demonstrate that charge, as opposed to spin, degrees of freedom
are responsible for glassy ordering of the 2DES near the MIT.

Previous studies carried out at $B=0$ employed a combination of
transport and low-frequency resistance noise
measurements~\cite{SBPRL02,JJPRL02} to probe the glassy behavior.
The glass transition was manifested by a sudden and dramatic
slowing down of the electron dynamics and by an abrupt change to
the sort of statistics characteristic of complicated multistate
systems, consistent with the hierarchical picture of glassy
dynamics.  It was also found~\cite{SBPRL02} that the glass
transition occurs in the metallic phase, {\it i.e.} at an electron
density $n_g>n_c$, where $n_c$ is the critical density for the MIT
determined from the vanishing of activation energy in the
insulating regime~\cite{Pudalov,Shashkin}.  The intermediate
metallic glass (MG) phase is of considerable width in strongly
disordered samples ($(n_g-n_c)/n_c\approx 0.5$ \cite{SBPRL02}),
whereas in devices with low disorder $n_g$ is only a few percent
higher than $n_c$ \cite{JJPRL02}, in agreement with theory
\cite{Darko}.  In this work, we investigate the glass transition
in these same high peak mobility (low disorder) devices using
noise spectroscopy in a parallel $B$.  We find that all the
signatures of the glass transition in parallel $B$ are
qualitatively the same as in $B=0$, even when the electrons are
spin polarized at high $B$.  We construct a phase diagram in the
$(n_s,B)$ plane ($n_s$ is the electron density), and find that the
MG phase is broadened by a parallel $B$. The temperature ($T$)
dependence of the conductivity $\sigma$ in the MG at low $B$ is
similar to the one first observed in the MG phase of highly
disordered devices at $B=0$~\cite{SBPRL02}, and it is consistent
with theory \cite{Denis}.  We show that, at low $B$, it also
provides evidence for a quantum phase transition (QPT) at $n_c
(B)$.

The experiment was performed on n-channel Si
metal-oxide-semiconductor field-effect transistors (MOSFETs) with
the peak mobility $\mu\approx 2.5$~m$^2$/Vs at 4.2~K, fabricated
in a Hall bar geometry with Al gates, and oxide thickness
$d_{ox}=147$~nm \cite{heemskerksamples}.  The resistance $R$ was
measured using a standard four-probe ac technique (typically
$2.7$~Hz) in the Ohmic regime. A precision DC voltage standard
(EDC MV116J) was used to apply the gate voltage, which controls
$n_s$. We note that $n_s$ was always varied at a relatively high
$T\approx 2$~K. Contact resistances and their influence on noise
measurements were minimized by using a split-gate geometry, which
allows one to maintain high $n_{s}$ ($\approx10^{12}$~cm$^{-2}$)
in the contact region while allowing an independent control of
$n_s$ of the 2D system under investigation in the central part of
the sample ($120\times 50~\mu$m$^2$).  The samples and the
measurement technique have been described in more detail elsewhere
\cite{JJPRL02}.

For each $n_s$ and $B$, $R$ was measured as a function of time $t$
at $T=0.24$~K, although measurements at higher $T$ were also
performed at several selected $B$.  At $B=0$, the temperature
coefficient of the time-averaged resistivity
$d\langle\rho\rangle/dT=0$ at $n_{s}^{\ast}\approx 9.7\times
10^{10}$cm$^{-2}$, similar to what was obtained on the previous
cooldown of the same sample \cite{JJPRL02}.  At a fixed $T$ and in
the range of $n_s$ under investigation, $\langle\rho\rangle$
exhibits a dramatic increase with $B$, followed by a much weaker
dependence (``saturation'') at higher fields ($B>2-4$~T).  This
large positive magnetoresistance at low $B$ has been observed and
studied extensively in many 2D systems \cite{SAK2000}, including
other samples from the same source \cite{heemskerksamples} as
ours.  In the saturation region, it has been shown
\cite{polarization} that the 2DES is spin polarized.

\begin{figure}
\centerline{\epsfig{file=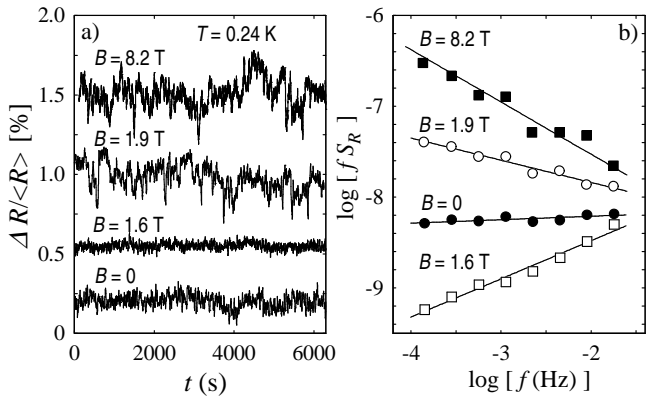,width=8.5cm,clip=}} \caption{
(a)$\Delta R(t)/\langle R\rangle$, and (b) the corresponding power
spectra $S_{R}(f)$, at several $B$ for
$n_s=11.9\times10^{10}$cm$^{-2}$. In (a) traces are shifted for
clarity.  In (b) $S_R(f)$ are averaged over octaves and multiplied
by $f$, so that $1/f$ spectrum is horizontal on this scale.  Solid
lines are linear least-squares fits with the slopes $\alpha=$
1.58, 1.24, 0.96, 0.58 (from top). \label{raw}}
\end{figure}

Figures~\ref{raw}(a) and \ref{raw}(b) show the time series of the relative
changes in resistance $\Delta R(t)/\langle R\rangle$ and the corresponding
power spectra $S_R(f)$, respectively, for a fixed $n_s$ and several $B$.  It
is obvious that $B$ has a strong effect on both the
%
%%%%%%%%%%%%%%% figure 1 %%%%%%%%%%%%%%%%%%%%%%%%%%%%%%%%%%%%%%%%%%%%
%%%%%%%%%%%%%%% figure 1 %%%%%%%%%%%%%%%%%%%%%%%%%%%%%%%%%%%%%%%%%%%%
%
amplitude and the character of the noise, as discussed in detail below.  In
order to compare the noise magnitudes under different conditions, the power
$S_R(f=1$~mHz) is taken as the measure of noise.  It is determined from the
fits of the octave-averaged spectra to the form $1/f^{\alpha}$ for
$10^{-4}<f<0.07$~Hz [solid lines in Fig. \ref{raw}(b)].
In addition, we have also analyzed the so-called second spectrum $S_2(f_2,f)$,
which is the power spectrum of the fluctuations of $S_{R}(f)$ with time
\cite{Weissman}.  $S_2(f_2,f)$ provides a direct probe of correlations between
fluctuators: it is white (independent of $f_2$) for uncorrelated, and
$S_2\propto 1/f_{2}^{1-\beta}$ for interacting fluctuators \cite{Weissman}.
At $B=0$, the glass transition in Si MOSFETs was manifested by a sudden and
dramatic increase of $S_{R}$, a rapid rise of $\alpha$ from $\approx 1$ to
$\approx 1.8$ \cite{SBPRL02,JJPRL02}, and a change of the exponent $(1-\beta)$
from a white (zero) to a nonwhite (nonzero) value \cite{JJPRL02}.  We adopt
similar criteria for the glass transition in $B$.

\begin{figure}
\centerline{\epsfig{file=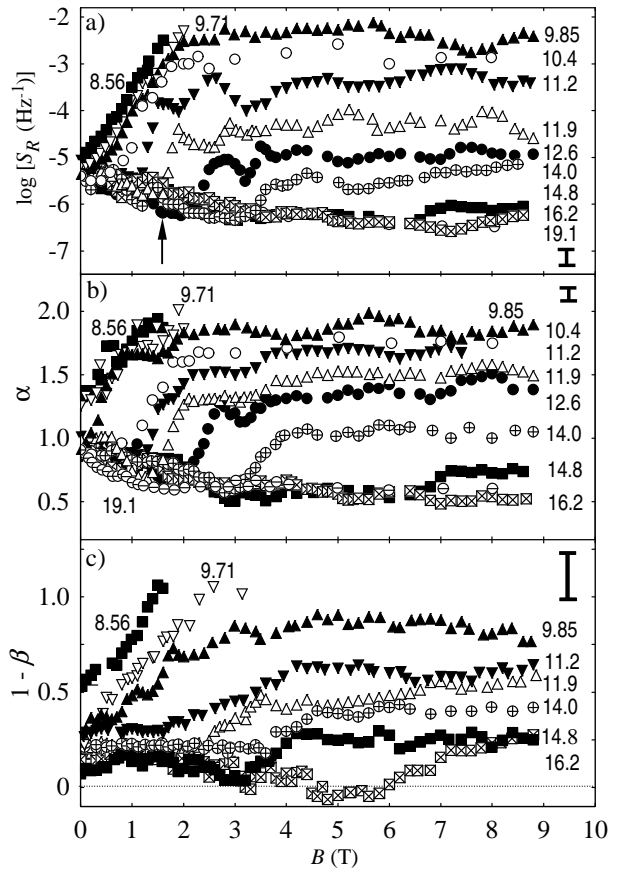,width=8cm,clip=}} \caption{(a)
$S_R(f=1$~mHz), (b) $\alpha$, and (c) $(1-\beta)$  {\it vs.} $B$
for $n_s(10^{10}$cm$^{-2})$ shown on the plots; $T=0.24$~K.  Other
data have been omitted for clarity. $S_{R}(f)$ has been corrected
for the white background noise. The arrow in (a) shows $B_g$ for
$n_s(10^{10}$cm$^{-2})=11.9$. The error bars on the right show
maximum standard deviations of the data, which, for clarity, were
plotted after performing a simple 3-point average. The origin of
the fluctuations of $S_R(B)$ for some $n_s$ at $B\gtrsim B_g$
(\textit{e. g.} for $n_s(10^{10}$cm$^{-2})=11.2$) is not
understood at this time. \label{vsb}}
\end{figure}

Figures \ref{vsb}(a), \ref{vsb}(b), and \ref{vsb}(c) present the
$B$-field dependences of $S_R$, $\alpha$, and $(1-\beta)$,
respectively, for several $n_s$.  For each $n_s$ and $B$,
$S_2(f_2,f)$ was calculated \cite{s2} for three different octaves
$f$: 2-4 mHz, 4-8 mHz, and 8-16 mHz.  In order to reduce the
uncertainty in $(1-\beta)$, the exponent shown in Fig. \ref{vsb}(c)
represents the average $(1-\beta)$ obtained from $S_2$ in those
three octaves.
%
%%%%%%%%%%%%%%% figure 2 %%%%%%%%%%%%%%%%%%%%%%%%%%%%%%%%%%%%%%%%%%%%
%%%%%%%%%%%%%%% figure 2 %%%%%%%%%%%%%%%%%%%%%%%%%%%%%%%%%%%%%%%%%%%%

At the highest $n_{s}$, in the metallic phase, ``pure'' $1/f$
noise ($\alpha\simeq 1$) resulting from uncorrelated
[$(1-\beta)\approx 0$] fluctuators is observed in $B=0$
\cite{SBPRL02,JJPRL02}.  A decreasing $S_{R}$ with $B$ in
Fig.~\ref{vsb}(a) shows that parallel $B$ suppresses such noise,
the behavior that is already apparent from the raw data
[\textit{e.g.} bottom two traces in Fig. \ref{raw}(a)].  In
addition, the exponent $\alpha$ is reduced to $\sim 0.5$ [Fig.
\ref{vsb}(b)]. The dependence of both $S_R$ and $\alpha$ on $B$
becomes weaker for $B\gtrsim 3-4$~T. We note that noise in the
metallic phase does not depend on $n_s$, similar to the $B=0$ case
\cite{SBPRL02,JJPRL02}. Even though the origin of this noise is
not known at the moment, these data should provide a valuable
contribution towards understanding the nature of the metallic
phase.  Pure $1/f$ noise in $B=0$ has been observed recently also
in a 2D hole system in GaAs near the apparent MIT, on the metallic
side \cite{LHote}.

\begin{figure}
\centerline{\epsfig{file=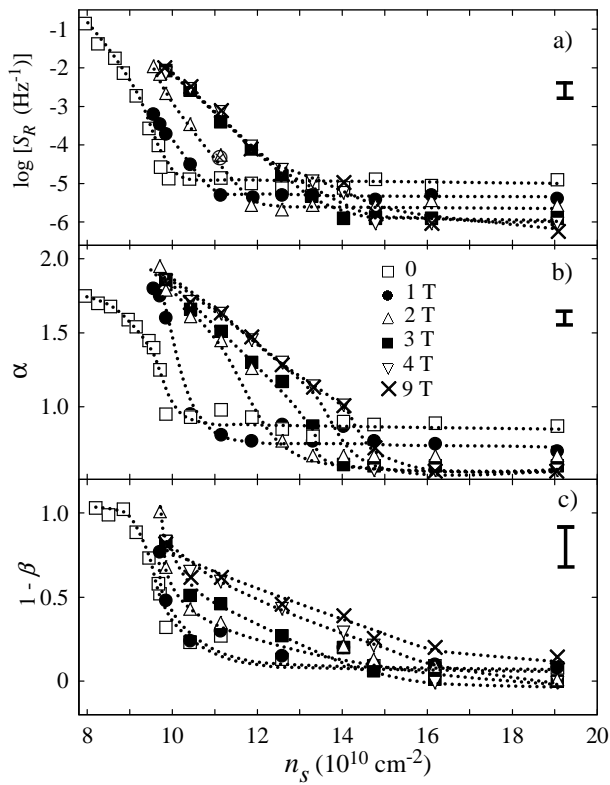,width=8cm,clip=}} \caption{(a)
$S_R(f=1$~mHz), (b) $\alpha$, and (c) $(1-\beta)$ \textit{vs.}
$n_s$ for several $B$ shown on the plot.  The same symbols are
used in all three panels. Dotted lines
guide the eye.  The maximum error bars are shown in the upper
right. \label{vsn}}
\end{figure}

At lower $n_{s}$, the noise at low $B$ behaves as described above.
However, for each $n_s$, there is now a well-defined field $B_g$
[Fig. \ref{vsb}(a)] where, after the initial decrease with $B$, $S_R$
begins to increase dramatically, accompanied by a rapid rise of
$\alpha$ and an increase of $(1-\beta)$ to nonwhite values,
indicative of the onset of strong correlations. This striking
change of both the magnitude and character of noise within a
narrow range of fields is obvious even from the raw data
(\textit{e.g.} middle two traces in Fig. \ref{raw}).  In analogy
with the $B=0$ case, we identify $B_g$ as the field where glass
transition occurs at a given $n_s$.  At even higher fields
($B\gtrsim 4$~T), $S_R$, $\alpha$, and $(1-\beta)$ no longer
depend on $B$, consistent with the fact that the electrons are
spin polarized.  Nevertheless, Fig. \ref{vsb} shows that the
strong dependence of noise on $n_s$ is still present for a given
$B$ in this regime.  The dependence of $S_R$, $\alpha$, and
$(1-\beta)$ on $n_s$ is plotted explicitly in Fig. \ref{vsn} for
several $B$.  The density where $S_R$ begins to increase
dramatically, accompanied by changes in $\alpha$ and $(1-\beta)$,
is identified as the glass transition density $n_g$ for a given
$B$.
%
%%%%%%%%%%%%%%% figure 3 %%%%%%%%%%%%%%%%%%%%%%%%%%%%%%%%%%%%%%%%%%%%
%%%%%%%%%%%%%%%  figure 3 %%%%%%%%%%%%%%%%%%%%%%%%%%%%%%%%%%%%%%%%%%%%

\begin{figure}
\centerline{\epsfig{file=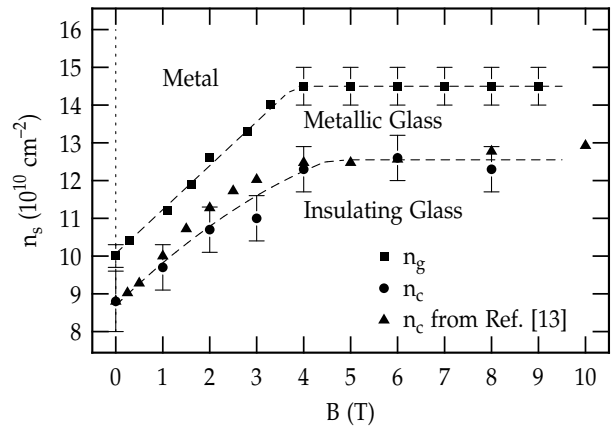,width=8cm,clip=}}
\caption{$T=0$ phase diagram.  The dashed lines guide the eye.
Squares: boundary between the metallic and MG phases; other symbols:
boundary between the MG and insulating glass phases. The data from
Ref. \cite{Shashkin} have been shifted up by $0.85\times
10^{10}$cm$^{-2}$ to make the $n_c(B=0)$ values coincide.
\label{phase}}
\end{figure}

The values of $B_g(n_s)$ and $n_g(B)$ determined in this way have
been used to construct a phase diagram in the $(n_s,B)$ plane, as
shown in Fig.~\ref{phase}.  The square symbols designate the
boundary of the glassy phase, and clearly show an increase of
$n_g$ with $B$, followed by a saturation at higher fields.  It is
%
%%%%%%%%%%%%%%% figure 4 %%%%%%%%%%%%%%%%%%%%%%%%%%%%%%%%%%%%%%%%%%%%
%%%%%%%%%%%%%%%  figure 4 %%%%%%%%%%%%%%%%%%%%%%%%%%%%%%%%%%%%%%%%%%%%
interesting to compare the behavior of $n_g(B)$ with $n_c(B)$.  In
Ref. \cite{Shashkin}, where samples almost identical to ours were
used, $n_c(B)$ was determined based on both a vanishing activation
energy and a vanishing nonlinearity of current-voltage
characteristics when extrapolated from the insulating phase.  We
have used the activation energy method but the range of accessible
$T$ and $n_s$ in our ac measurements was smaller compared to the
dc technique and dilution refrigerator $T$ (down to 30 mK) used in
Ref. \cite{Shashkin}.  We find that the data that we have available in
the insulating regime are best described by
$\langle\sigma\rangle\propto\exp [-(T_{0}/T)^{n}]$ with $n=1/2$,
which corresponds to variable-range hopping with a Coulomb gap
\cite{ES}.  The extrapolation of $T_0(n_s)$ to zero, where only
$n_s$ with $T_0\gtrsim 0.5$~K were used,
was employed to determine $n_c$ shown in Fig. \ref{phase}.  Strictly
speaking, the rather limited range of data does not allow one to
make an accurate distinction between different forms of activated
$\langle\sigma(T)\rangle$. This experimental uncertainty is reflected
in rather large error bars for $n_c$ shown in Fig. \ref{phase}.
Nevertheless, the agreement between our results and those obtained
in Ref. \cite{Shashkin} is remarkably good (Fig. \ref{phase}).

\begin{figure}
\centerline{\epsfig{file=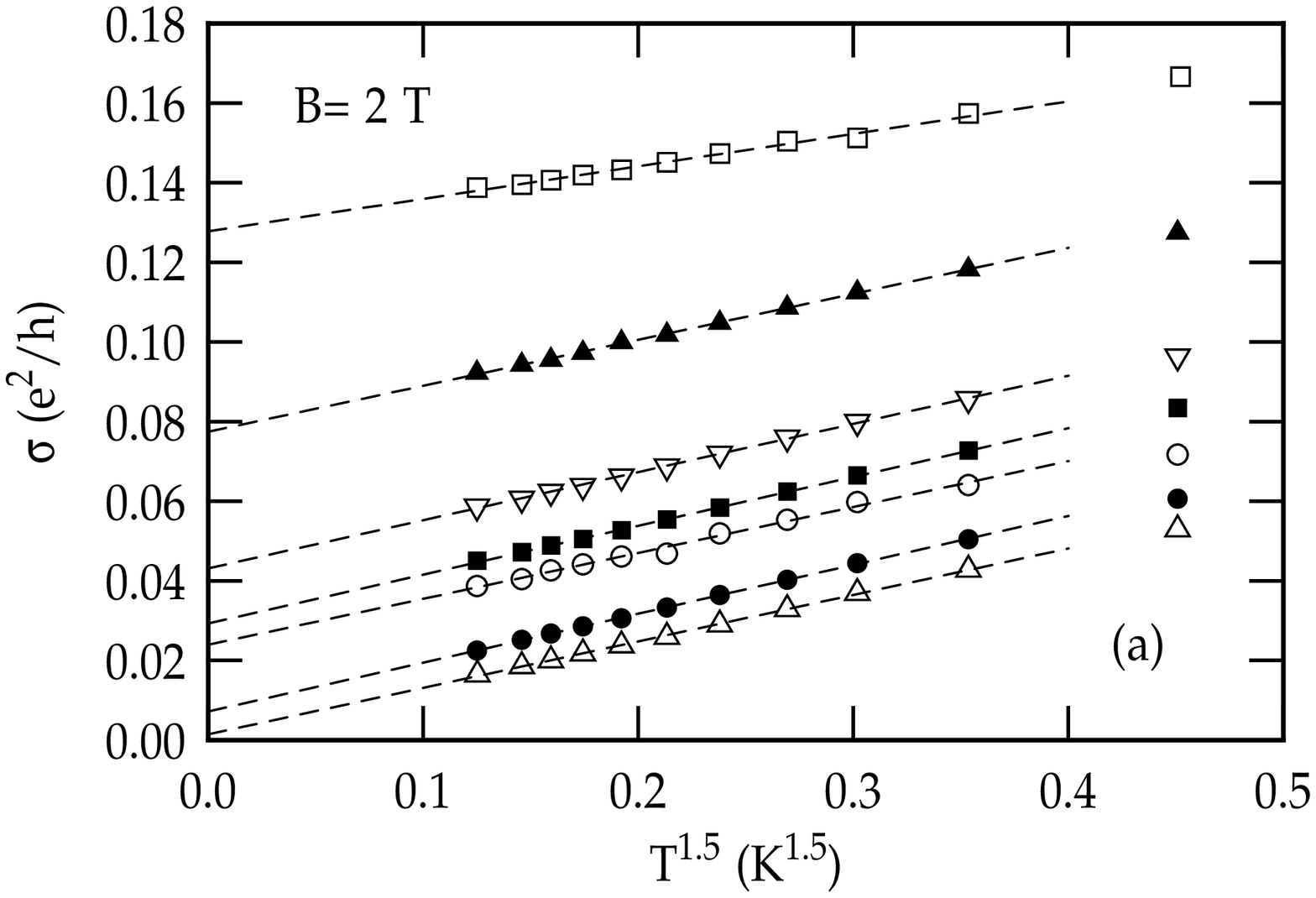,width=8cm,clip=}}
\centerline{\epsfig{file=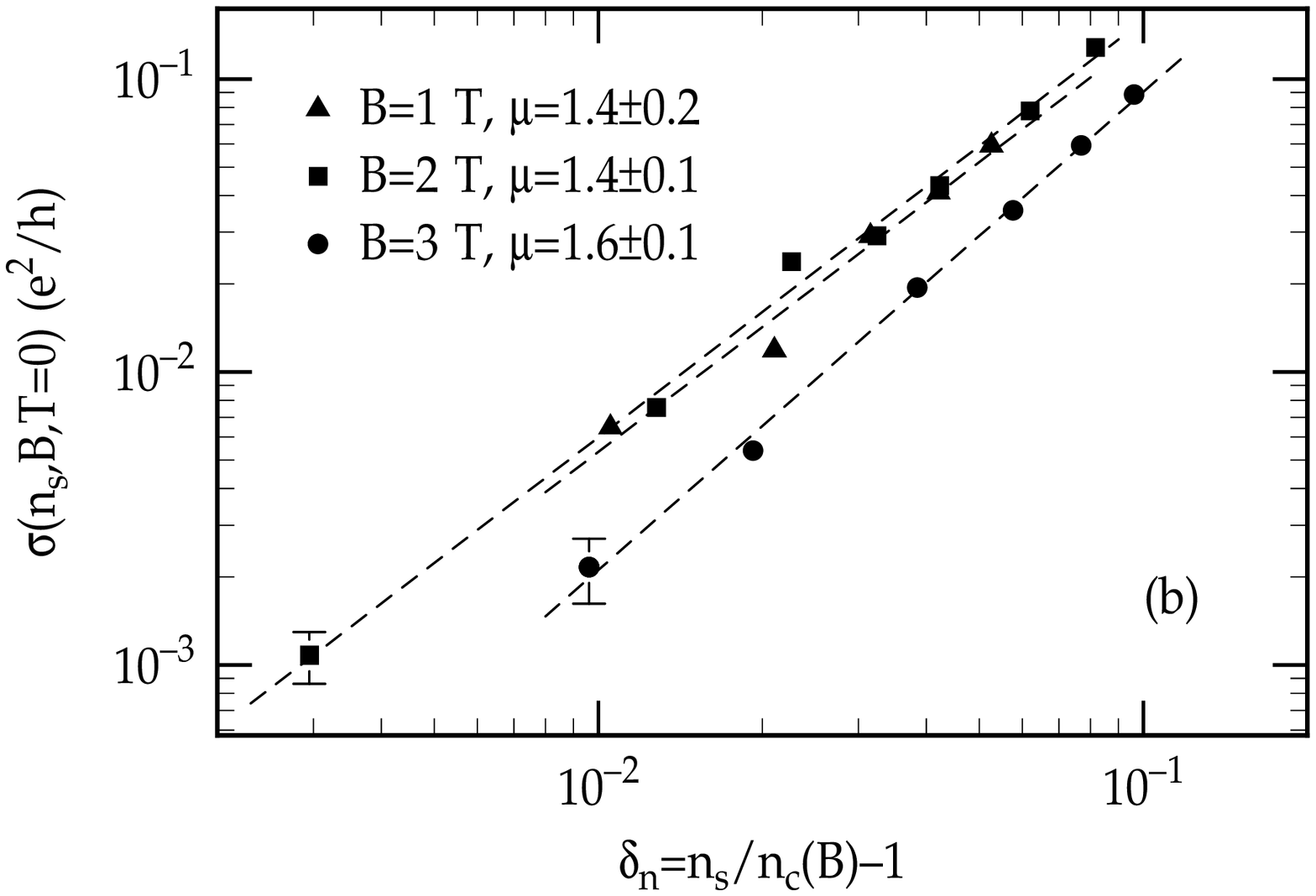,width=8cm,clip=}}
\caption{(a) $\langle\sigma(T)\rangle$ in the MG phase for
$n_s(10^{10}$cm$^{-2})=11.9, 11.6, 11.3, 11.2, 11.0, 10.9, 10.7$
from top; $B=2$~T; $n_c(B=2$~T$)=10.67\times10^{10}$cm$^{-2}$.  (b)
$\langle\sigma(T=0)\rangle\propto\delta_{n}^{\mu}$.
 \label{tempdep}}
\end{figure}

Furthermore, it was possible for the first time to determine $n_c$ in a
magnetic field by studying $\langle\sigma(T)\rangle$ on the
\emph{metallic} side of the MIT.  In particular, in a relatively narrow
range of $n_s$ for $B=1, 2, 3$~T, the data are best described
by the metallic power-law behavior
$\langle\sigma(n_s,B,T)\rangle=\langle\sigma(n_s,B,T=0)\rangle
+b(n_s,B)T^{1.5}$ [Fig. \ref{tempdep}(a)], similar to what was
observed in the MG phase of highly disordered samples at $B=0$
\cite{SBPRL02}. The extrapolated $T=0$ conductivities go to zero
at $n_c(B)$ in a power-law fashion
$\langle\sigma(n_s,B,T=0)\rangle\propto\delta_{n}^{\mu}$ with
$\mu\sim 1.5$ [see Fig. \ref{tempdep}(b);
$\delta_{n}=n_s/n_c(B)-1$], in agreement with theoretical
expectations near a QPT \cite{Belitz}. At $B=0$, there is some
evidence \cite{Fletcher} of similar behavior with $\mu\sim 1-1.5$,
obtained from $\sigma(T=0)$ in the metallic phase of different Si
MOSFETs. Considering that $n_c(B)$ for $B=1,2,3$~T shown in
Fig. \ref{phase} have been obtained by approaching the MIT from
the \emph{metallic} side, the agreement between our results and
those of Ref. \cite{Shashkin} is
%
%%%%%%%%%%%%%%% figure 5 %%%%%%%%%%%%%%%%%%%%%%%%%%%%%%%%%%%%%%%%%%%%
%%%%%%%%%%%%%%%  figure 5 %%%%%%%%%%%%%%%%%%%%%%%%%%%%%%%%%%%%%%%%%%%%
even more remarkable.

The phase diagram in Fig. \ref{phase} makes it clear why the
metallic $T^{3/2}$ temperature dependence of conductivity [Fig.
\ref{tempdep}(a)] is observed in such a narrow range of $n_s$: it
is characteristic of the metallic glass phase, even in $B$ of up
to 3~T. At $B=0$, the MG phase is very narrow and the
$T^{3/2}$ correction is, therefore, difficult to observe in these
samples, in contrast to highly disordered ones \cite{SBPRL02}.  A
parallel $B$ increases the width of the MG phase (Fig.
\ref{phase}).  In addition, the range of $T$ where $T^{3/2}$ holds
increases with parallel $B$ (not shown).  The increase of $n_g$
and $n_c$, and the broadening of the MG phase with $B$ can be
understood to result from the suppression of screening by a
parallel $B$ \cite{Bscreening}, which increases the effective
disorder. This, in turn, favors glassiness, consistent with
theoretical expectations \cite{Darko}.  It is also interesting to
note that the prefactor $b(n_s,B)$ of the $T^{3/2}$ correction
[slopes in Fig. \ref{tempdep}(a)] does not seem to depend on
$n_s$, in contrast to the strong $n_s$-dependence seen in highly
disordered (low-mobility) samples at $B=0$ \cite{SBPRL02}. Further
careful investigation is required in order to determine whether
this difference can be attributed to the effects of $B$
or to the effects of disorder.  Such a study, however, along with a
detailed analysis of $\langle\sigma(T)\rangle$ in the MG phase at high
fields, is beyond the scope of this paper.

In summary, the glass transition in high-mobility Si MOSFETs has
been found to persist in high parallel $B$, where the 2DES is spin
polarized.  These results demonstrate that charge, as opposed to
spin, degrees of freedom are responsible for glassy ordering of
the 2DES near the MIT, consistent with recent theory
\cite{Pastor,Denis,Darko}.

We are grateful to V. Dobrosavljevi\'c for useful discussions.  This work was
supported by NSF grant No. DMR-0071668, and NHMFL through NSF Cooperative
Agreement No. DMR-0084173.

%\bibliography{highmuB}

\begin{thebibliography}{22}
\expandafter\ifx\csname natexlab\endcsname\relax\def\natexlab#1{#1}\fi
\expandafter\ifx\csname bibnamefont\endcsname\relax
  \def\bibnamefont#1{#1}\fi
\expandafter\ifx\csname bibfnamefont\endcsname\relax
  \def\bibfnamefont#1{#1}\fi
\expandafter\ifx\csname citenamefont\endcsname\relax
  \def\citenamefont#1{#1}\fi
\expandafter\ifx\csname url\endcsname\relax
  \def\url#1{\texttt{#1}}\fi
\expandafter\ifx\csname urlprefix\endcsname\relax\def\urlprefix{URL }\fi
\providecommand{\bibinfo}[2]{#2}
\providecommand{\eprint}[2][]{\url{#2}}

\bibitem[{SAK()}]{SAK2000}
\bibinfo{note}{E. Abrahams, S. V. Kravchenko, and M. P. Sarachik, Rev. Mod.
  Phys. {\bf 73}, 251 (2000), and references therein}.

\bibitem[{SBP()}]{SBPRL02}
\bibinfo{note}{S. Bogdanovich and D. Popovi\'c, Phys. Rev. Lett. {\bf 88},
  236401 (2002)}.

\bibitem[{JJP()}]{JJPRL02}
\bibinfo{note}{J. Jaroszy\'nski, D. Popovi\'c, and T. M. Klapwijk, Phys. Rev.
  Lett. {\bf 89}, 276401 (2002)}.

\bibitem[{Tha()}]{Thakur}
\bibinfo{note}{J. S. Thakur and D. Neilson, Phys. Rev. B {\bf 54} 7674 (1996);
  {\bf 59}, R5280 (1999)}.

\bibitem[{Pas()}]{Pastor}
\bibinfo{note}{A. A. Pastor and V. Dobrosavljevi\'c, Phys. Rev. Lett. {\bf 83},
  4642 (1999)}.

\bibitem[{Sud()}]{Sudip}
\bibinfo{note}{S. Chakravarty {\em et al.}, Philos. Mag. B {\bf 79}, 859
  (1999)}.

\bibitem[{Sac()}]{Sachdev}
\bibinfo{note}{S. Sachdev, Pramana {\bf 58}, 285 (2002), cond-mat/0109309;
  Philos. Trans. R. Soc. London A {\bf 356}, 173 (1998)}.

\bibitem[{Den()}]{Denis}
\bibinfo{note}{D. Dalidovich \emph{et al.}, Phys. Rev. B {\bf 66}, 081107
  (2002)}.

\bibitem[{Dar()}]{Darko}
\bibinfo{note}{V. Dobrosavljevi\'c, D. Tanaskovi\'c, and A. A. Pastor, Phys.
  Rev. Lett. {\bf 90}, 016402 (2003)}.

\bibitem[{Bin()}]{Binder}
\bibinfo{note}{K. Binder {\em et al.}, Rev. Mod. Phys. {\bf 58}, 801 (1986)}.

\bibitem[{com()}]{comment1}
\bibinfo{note}{As usual, parallel field is used in order to avoid complications
  due to the orbital motion of electrons.}

\bibitem[{Pud()}]{Pudalov}
\bibinfo{note}{V. M. Pudalov {\em et al.}, Phys. Rev. Lett. {\bf 70}, 1866
  (1993)}.

\bibitem[{Sha()}]{Shashkin}
\bibinfo{note}{A. A. Shashkin {\em et al.}, Phys. Rev. Lett. {\bf 87}, 266402
  (2001)}.

\bibitem[{hee()}]{heemskerksamples}
\bibinfo{note}{R. Heemskerk Ph.D. thesis, University of Groningen, The
  Netherlands, 1998 (unpublished); R. Heemskerk and T. M. Klapwijk, Phys. Rev.
  B {\bf 58}, R1754 (1998)}.

\bibitem[{pol()}]{polarization}
\bibinfo{note}{T. Okamoto \emph{et al.}, Phys. Rev. Lett. \textbf{82}, 3875
  (1999); S. A. Vitkalov \emph{et al.}, Phys. Rev. Lett. \textbf{85}, 2164
  (2000); E. Tutuc \emph{et al.}, Phys. Rev. Lett. \textbf{86}, 2858 (2001)}.

\bibitem[{Wei()}]{Weissman}
\bibinfo{note}{M. B. Weissman, Rev. Mod. Phys. {\bf 60}, 537 (1988); Rev. Mod.
  Phys. {bf 65}, 829 (1993); M. B. Weissman {\em et al.}, J. Magn. Magn. Mater.
  {\bf 114}, 87 (1992), and references therein.}

\bibitem[{s2()}]{s2}
\bibinfo{note}{G. T. Seidler and S.A. Solin, Phys. Rev. B \textbf{53}, 9753
  (1996); K. M. Abkemeier, Phys. Rev. B \textbf{55}, 7005 (1997)}.

\bibitem[{LHo()}]{LHote}
\bibinfo{note}{R. Leturcq {\em et al.}, Phys. Rev. Lett. {\bf 90}, 076402
  (2003)}.

\bibitem[{ES()}]{ES}
\bibinfo{note}{A. L. Efros and B. I. Shklovskii, J. Phys. C {\bf 8}, L49
  (1975)}.

\bibitem[{Bel()}]{Belitz}
\bibinfo{note}{D. Belitz {\em et al.}, Rev. Mod. Phys. \textbf{66}, 261
  (1994)}.

\bibitem[{Fle()}]{Fletcher}
\bibinfo{note}{R. Fletcher {\em et al.}, Semicond. Sci. Tech. \textbf{16}, 386
  (2001)}.

\bibitem[{Bsc()}]{Bscreening}
\bibinfo{note}{V. T. Dolgopolov and A. Gold, JETP Lett. \textbf{71}, 27 (2000);
  I. F. Herbut, Phys. Rev. B \textbf{63}, 113102 (2001)}.

\end{thebibliography}

\newcommand{\noopsort}[1]{} \newcommand{\printfirst}[2]{#1}
  \newcommand{\singleletter}[1]{#1} \newcommand{\switchargs}[2]{#2#1}

\end{document}